\documentclass[twocolumn,aps,prd,showpacs,amsmath,amssymb]{revtex4-1}
\usepackage[]{latexsym}
\usepackage[english]{babel}
\usepackage{graphicx}
\usepackage{hyperref}
\usepackage{color}

\newcommand{\abs}[1]{\vert{#1}\vert}

\begin{document}

\title{Cosmographic reconstruction of $f(\mathcal{T})$ cosmology}

\author{Alejandro Aviles}
\email{aviles@ciencias.unam.mx}
\affiliation{Depto. de F\'isica, Instituto Nacional de Investigaciones Nucleares, M\'exico, Mexico}
\affiliation{Instituto de Ciencias Nucleares, Universidad Nacional Aut\'onoma de M\'exico, AP 70543, M\'exico, DF 04510, Mexico.}

\author{Alessandro Bravetti}
\email{bravetti@icranet.org}
\affiliation{Dip.  di Fisica and ICRA, "Sapienza" Universit\`a di Roma,
            Piazzale Aldo Moro 5, I-00185, Roma, Italy}
\affiliation{Instituto de Ciencias Nucleares, Universidad Nacional Aut\'onoma de M\'exico, AP 70543, M\'exico, DF 04510, Mexico.}

\author{Salvatore Capozziello}
\email{capozzie@na.infn.it}
\affiliation{Dip. di  Fisica, Universit\`a di Napoli "Federico II",
             Via Cinthia, I-80126, Napoli, Italy}
\affiliation{INFN Sez. di Napoli, Compl. Univ. Monte S. Angelo Ed. N
Via Cinthia, I- 80126 Napoli, Italy.}

\author{Orlando Luongo}
\email{orlando.luongo@na.infn.it}
\affiliation{Instituto de Ciencias Nucleares, Universidad Nacional Aut\'onoma de M\'exico, AP 70543, M\'exico, DF 04510, Mexico;}
\affiliation{Dip. di Fisica, Universit\`a di Napoli "Federico II",
             Via Cinthia, I-80126, Napoli, Italy}
\affiliation{INFN Sez. di Napoli, Compl. Univ. Monte S. Angelo Ed. N
Via Cinthia, I- 80126, Napoli, Italy.}

\date{\today}

\begin{abstract}
A cosmographic reconstruction of $f(\mathcal T)$ models is here revised in a model independent way by fixing observational bounds on the most relevant terms of the $f(\mathcal T)$ Taylor expansion.
We relate the $f(\mathcal T)$ models  and their derivatives to  the cosmographic parameters and  then adopt a Monte Carlo analysis.  The experimental bounds  are thus independent of the choice of  a particular $f(\mathcal T)$ model. The advantage of such an analysis lies on constraining the dynamics of the universe by reconstructing the form of $f(\mathcal T)$, without any further assumptions apart from the validity of the cosmological principle and the analyticity   of the $f(\mathcal T)$ function.
The main result is to fix model independent cosmographic constraints on the functional form of $f(\mathcal T)$ which are compatible with the theoretical predictions.
Furthermore, we infer a phenomenological expression for $f(\mathcal T)$, compatible with the current cosmographic bounds and  show that small deviations are expected from a constant $f(\mathcal T)$ term, indicating that the equation of state of dark energy could slightly evolve from the one of the $\Lambda$CDM model.
\end{abstract}


\maketitle

\section{Introduction}

Current astrophysical observations predict the existence of  further fluids, in addition to the standard pressureless matter terms, which allow  the universe to undergo a positive late time acceleration \cite{SNeIa1,SNeIa2,SNeIa3}. Unfortunately, the physical origin of such exotic  fluids is  a big puzzle   in  modern cosmology \cite{JorgeSmoot}. Particularly, an intense debate arose in order to understand which kind of fluid can be responsible for this apparent  cosmic speed up \cite{copeland}. Even though its physical nature is completely unclear, it is well understood that it shows  a negative equation of state (EoS) \cite{v1,v2,v3}. Such an  EoS is able to accelerate the universe because it provides an antigravitational effect, which counterbalances the attraction of gravity acting on standard matter \cite{mioprocedd,dopocopeland}. The fluid, which drives the observed acceleration, is usually referred to as dark energy (DE) and fills more than the 70$\%$ of the whole universe energy budget \cite{bston}.

One of the most viable choices for DE is to assume the existence of a vacuum energy cosmological constant $\Lambda$ \cite{lambda1,lambda2}. The corresponding  $\Lambda$CDM model, is characterized by the presence of baryon and cold dark matter components intertwined with the cosmological constant density, $\Omega_\Lambda$, which actually represents the natural candidate for depicting the DE effects \cite{lambdax}. The model excellently passes all the experimental tests and entered the modern cosmological convictions, becoming the new standard cosmological model.
Even though $\Lambda$CDM seems to be experimentally favorite in explaining the universe dynamics, it is plagued by two profound issues, i.e. the problems of coincidence and fine-tuning (for further details see \cite{miao}). Thus, other alternative explanations have been carried on during the last decades \cite{mine,mine2}. The common property of all these approaches is that the DE effects can be associated to an evolving EoS \cite{u1,u2,u3}; among the wide number of different paradigms, a relevant role is played by theories that claim to extend general relativity (GR). The underlying philosophy of  extended theories of gravity is that GR should be seen as a particular  case of a more general effective theory coming from fundamental principles  \cite{rev1,rev2,rev3,rev4,rev5,altro,altrissimo1,altrissimo2,altrissimo3}.
The basic idea lies on the fact that the standard Einstein-Hilbert action is modified by additional degrees of freedom, spanning from further curvature invariant corrections,  to scalar fields and Lorentz violating terms.

Here, we limit our attention to investigate the so called $f(\mathcal T)$ theories, which represent a class of models that take into account the effects due to the torsion \cite{Teleparallelism, storny,bamba}. The main advance of $f(\mathcal T)$ gravity deals with the fact that the Ricci scalar curvature, $\mathcal R$,
is assumed to be $0$. Under this hypothesis, the $f(\mathcal T)$ models can be seen as extensions of the so-called Teleparallel gravity \cite{Teleparallelism}. The property of a vanishing $\mathcal{R}$ is motivated by the theoretical inflationary phases in the early stages of the universe evolution. Thus, one can extend the Lagrangian density by adding a torsion function, i.e. $f(\mathcal T)$.

To infer the cosmological equations, we can consider an orthonormal basis $e_A (x^{\mu})$ for the tangent space at each point $x^{\mu}$ on a generic manifold, with $A$ running over $0, 1, 2, 3$. Therefore, the metric tensor is obtained from the dual basis $e^A (x^{\mu})$ as
\begin{equation}\label{n}
g_{\mu\nu}=\eta_{A B} e^A_\mu e^B_\nu\,,
\end{equation}

\noindent where $\mu$ and $\nu$ are the coordinate indices on the manifold running over $0, 1, 2, 3$ and $e^A_\mu$ are the coordinates of the dual basis with respect to a coordinate basis on the cotangent space. It is prominent to redefine the torsion $\mathcal T^\rho_{\verb| |\mu\nu}$ in terms of the quoted quantities, obtaining

\begin{equation}
\mathcal T^\rho_{\verb| |\mu\nu} \equiv e^\rho_A
\left( \partial_\mu e^A_\nu - \partial_\nu e^A_\mu \right)\,,
\label{eq:IXA-2.2}
\end{equation}

\noindent and the so-called contorsion $K^{\mu\nu}_{\verb|  |\rho}$, that reads

\begin{equation}
K^{\mu\nu}_{\verb|  |\rho}
\equiv
-\frac{1}{2}
\left(\mathcal T^{\mu\nu}_{\verb|  |\rho} - \mathcal T^{\nu \mu}_{\verb|  |\rho} -
\mathcal T_\rho^{\verb| |\mu\nu}\right)\,.
\label{eq:IXA-2.3}
\end{equation}

\noindent For the sake of simplicity, one can write down

\begin{equation}\label{eq:IXA-2.5}
S_\rho^{\verb| |\mu\nu} \equiv \frac{1}{2}
\left(K^{\mu\nu}_{\verb|  |\rho}+\delta^\mu_\rho \
\mathcal T^{\alpha \nu}_{\verb|  |\alpha}-\delta^\nu_\rho \
\mathcal T^{\alpha \mu}_{\verb|  |\alpha}\right)\,,
\end{equation}

\noindent that, together with Eq. ($\ref{eq:IXA-2.2}$), allows to get the torsion scalar $\mathcal T$ as

\begin{equation}
\mathcal T \equiv S_\rho^{\verb| |\mu\nu} \mathcal T^\rho_{\verb| |\mu\nu}\,.
\label{eq:IXA-2.4}
\end{equation}

Finally, the modified teleparallel action of a generic $f(\mathcal T)$ model with the matter Lagrangian ${\mathcal{L}}_{\mathrm{M}}$
is

\begin{equation}
{\mathcal A}=
\int d^4x \abs{e} \left[ \frac{f(\mathcal T)}{2{\kappa}^2}
+{\mathcal{L}}_{\mathrm{M}} \right]\,,
\label{eq:IXA-2.6}
\end{equation}

\noindent where $\abs{e}= \det \left(e^A_\mu \right)=\sqrt{-g}$.
By varying Eq. ($\ref{eq:IXA-2.6}$) with respect to the vierbein vector field $e_A^\mu$, we get the gravitational field equations

\begin{eqnarray}\label{eq:IXA-2.7}
\frac{{\kappa}^2}{2} e_A^\rho
{\mathcal T^{(\mathrm{M})}}_\rho^{\verb| |\nu}&=&\frac{1}{e} \partial_\mu \left( eS_A^{\verb| |\mu\nu} \right) f^{\prime}-e_A^\lambda \mathcal T^\rho_{\verb| |\mu \lambda} S_\rho^{\verb| |\nu\mu}
f^{\prime}\\
&+&S_A^{\verb| |\mu\nu} \partial_\mu \left(\mathcal T\right) f^{\prime\prime}
+\frac{1}{4} e_A^\nu f \,,\nonumber
\end{eqnarray}

\noindent where ${\mathcal T^{(\mathrm{M})}}_\rho^{\verb| |\nu}$ is the energy-momentum tensor, including all the possible fluids, such as matter, radiation, and so on. From Eq. ($\ref{eq:IXA-2.7}$), it is easy to note that the problem of determining the DE fluid is switched to  find out  a suitable form for $f(\mathcal T)$.

In addition, a compelling observational challenge is to reconstruct the cosmological bounds on $f(\mathcal T)$ and its derivatives, without postulating any $f(\mathcal T)$ \emph{a priori}. This permits to feature the universe evolution, without imposing any cosmological model from the beginning. One of the most remarkable approaches, lying on being model independent, is represented by  {\it cosmography }\cite{altrissimo1,io1,io2}.  Cosmography was initially proposed in order to relate the Taylor expansion of the luminosity distance in terms of the acceleration parameter. Recently it raised much interest to constrain the dynamics of the universe. The only remarkable assumption is the validity of the cosmological principle and the use of the corresponding Friedmann-Robertson-Walker metric (FRW). Thus, cosmography allows us to infer how much DE or alternative components are requested in order to obtain the late time acceleration of the universe. As discussed in \cite{io2}, the idea of cosmography is to expand the physical observables, such as the cosmological distances, the Hubble parameter, the pressure and so forth, into a Taylor series around the epoch $z=0$. Afterwards, one can relate the model free parameters which are under exam directly to such observables. In doing so, it is straightforward to show which models are compatible with the kinematics of the universe and which ones should be discarded as a consequence of not satisfying the basic demands of cosmography \cite{frcosmo,mariam}.

The aim of this paper is to derive general constraints on $f(\mathcal T)$ and its derivatives at our time in terms of the observable cosmographic quantities. This procedure extends the work in \cite{capozziellogiafatto} by inverting the equations relating the $f(\mathcal T)$ Taylor series to the cosmographic parameters, and obtaining more accurate constraints on $f(\mathcal T)$ models. This is due to the fact that we minimize the error propagation in the fitting procedure, by relating \emph{directly} the luminosity distance in terms of the $f(\mathcal T)$ Taylor series. We proceed with a Monte Carlo analysis, performed by modifying the available CosmoMC code.  The corresponding experimental limits on the observable $f(\mathcal T)$ allow us to propose an analytic function for $f(\mathcal T)$, satisfying the cosmographic requirements (see also \cite{altrissimo1}).

The paper is organized as follows. In Sec. \ref{cosmography} we evaluate the cosmographic quantities of interest and relate them in terms of  $f(\mathcal T)$ quantities. In Sec. \ref{constraints}, we discuss the experimental procedure used to fix constraints on $f(\mathcal T)$ and its derivatives at $z=0$ and present possible examples of $f(\mathcal T)$ gravity compatible with our results. Finally, in Sec. \ref{conclusions}, we draw conclusions and discuss the results. The detailed expressions for the luminosity distance in terms of $f(\mathcal T)$ function  are given in the Appendix \ref{appendix}.

\section{Cosmography of $f(\mathcal T)$ gravity}\label{cosmography}

Let us discuss now the cosmographic analysis that we are going to perform for $f(\mathcal T)$ gravity. First of all it is important to stress that  cosmography  very marginally  depends on the choice of a given cosmological model, because  the only  stringent assumption is that the universe is homogeneous and isotropic. This is alternative to almost all the cosmological tests which basically  fix  priors and constraints for  the model under consideration. This fact can give rise to  severe degeneracy problems  and does not allow to infer which model is really favored. Cosmography, on the contrary,  alleviates  degeneracy, because it bounds only cosmological quantities which do not strictly  depend on a model. Thus, by postulating the validity of the cosmological principle, we consider the FRW metric, i.e.
\begin{equation}\label{frw}
ds^2=dt^2-a(t)^2\left(dr^2+r^2\,d\Omega^2\right)\,,
\end{equation}
where we assume hereafter a spatially flat universe ($k=0$), with $d\Omega^2\equiv d\theta^2+\sin^2\theta \,d\phi^2$ and  expand the scale factor $a(t)$ in a Taylor series around the present epoch $t_0$. From the expansion of the scale factor, we define the quantities
\begin{eqnarray}\label{pinza}
\mathcal{H} \equiv \frac{1}{a} \frac{da}{dt}\,,\quad&\,& \quad q \equiv -\frac{1}{a\mathcal{H}^2} \frac{d^2a}{dt^2}\,,\nonumber\\
j  \equiv \frac{1}{a\mathcal{H}^3} \frac{d^3a}{dt^3}\,,\quad&\,& \quad s \equiv \frac{1}{a\mathcal{H}^4} \frac{d^4a}{dt^4}\,.
\end{eqnarray}
which are, by construction,  model independent quantities, and are referred to as the cosmographic set (CS). They are known  as the Hubble rate ($\mathcal{H}$), the acceleration parameter ($q$), the jerk parameter ($j$), the snap parameter ($s$).

Here, we want to relate the CS to the derivatives of $f(\mathcal T)$ in order to set constraints on   the model  without any {\it a priori} assumption on   the function $f(\mathcal T)$. In order to perform this, let us sketch the main steps that we are going to perform along this work.

We will use the modified Friedmann equations and the definition of the CS, to write $f(\mathcal T)$ and its derivatives with respect to $\mathcal T$ at present time as functions of the CS. Then, we will algebraically invert these relations and write the luminosity distance in terms of $f(\mathcal T)$ and its derivatives. Once the luminosity distance is rewritten in terms of  $f(\mathcal T)$ and its derivatives, it is possible to constrain such coefficients {\it directly}, without postulating any $f(\mathcal T)$ function \emph{a priori}. In addition, our procedure allows to alleviate the problem of the error propagation in the numerical analysis.

To derive the expressions for $\mathcal H$ in terms of the CS introduced in (\ref{pinza}), let us start with the modified Friedmann equations in the case of $f(\mathcal T)$ gravity, i.e.

\begin{eqnarray}
\mathcal{H}^2& =& \frac{1}{3} \left [ \rho+  \rho_{\mathcal T} \right
]\,, \label{eq1}\\
2 \dot{\mathcal{H}} + 3\mathcal{H}^2&=& - \frac{1}{3} (p+p_{\mathcal T })\,,\label{eq2}
\end{eqnarray}
where the standard matter energy density $\rho$ and pressure $p$ have their torsion scalar  counterpart $\rho_{\mathcal T}$ and $p_{\mathcal T}$,
\begin{equation}\label{rhoT}
    \rho_{\mathcal T}=\frac{1}{2}[2{\mathcal T}f'({\mathcal T})-f({\mathcal T})-{\mathcal T}/2],
\end{equation}
\begin{equation}\label{pT}
    p_{\mathcal T}=\frac{1}{2}[2\dot H(4{\mathcal T}f''({\mathcal T})+2f'({\mathcal T})-1)]-\rho_{\mathcal T}.
\end{equation}
are the torsion contributions to the energy density and pressure. Then, by using Eqs. (\ref{rhoT}) and (\ref{pT}), we can define
the effective torsion equation of state as
\begin{equation}\label{omegaeff}
    \omega_{{\mathcal T}}\equiv\frac{p_{\mathcal T}}{\rho_{\mathcal T}}=-1+\frac{4\dot H(4{\mathcal T}f''({\mathcal T})+2f'({\mathcal T})-1)}{4{\mathcal T}f'({\mathcal T})-2f({\mathcal T})-{\mathcal T}}.
\end{equation}
The quantities  $\rho_{\mathcal T}$ and $p_{\mathcal T}$ can be  related to the observed acceleration of the universe. Here we assume to work in the so-called Jordan frame, where the action (\ref{eq:IXA-2.6}) is not the usual Hilbert-Einstein action and standard fluid matter is minimally coupled to geometry. This choice can be justified as follows.

In our calculations, we prevent any measurement departure from changing frame because it is possible to demonstrate that, passing through different frames, the cosmographic series preserve its form  \cite{flanagan, deruelle}. Hence, no physical modifications are expected in transforming the system from the Einstein frame to the Jordan frame and vice-versa. Particular attention has been devoted to the cosmographic series. In the case of modified gravity, the corresponding changes has been accounted and investigated in \cite{gunzig,brown}. Afterwards, the problem has been considered in \cite{bamba}.
By means of this equivalence, physics in both  frames is identical. It implies that a system sets up in the Jordan frame is solved in the Einstein frame as long as it is transformed back to the Jordan frame and vice versa. The corresponding invariance in form turns out to be evident by assuming
$\rho_{\mathcal T}$ and $p_{\mathcal T}$ in the Friedmann Eqs(\ref{eq1}) and (\ref{eq2}).  In particular
differentiating the Friedmann equations with respect to $t$ and using (\ref{pinza}),  one can  write
\begin{eqnarray}
    \dot{\mathcal H}&=&-\mathcal H^2(1+q)\,,\label{dotH}\\
    \ddot{\mathcal H}&=&\mathcal H^3(j+3q+2)\,,\label{ddotH}\\
    \dddot{ \mathcal H}&=&\mathcal H^4[s-4j-3q(q+4)-6]\,,\label{dddotH}
\end{eqnarray}
which represent the relations between $\mathcal H$ and the CS. The dots denote the derivatives with respect to the cosmic time $t$.

For a FRW universe we have
\begin{eqnarray}\label{eq: constr}
\mathcal{T} &=& - 6\, \mathcal{H}^2\,,
\end{eqnarray}
so that, differentiating Eq. (\ref{eq: constr}) with respect to $t$, one gets
\begin{eqnarray}
  \dot{ \mathcal T} &=& -12\,\mathcal H\dot{\mathcal H}\,, \label{dif1T} \\
  \ddot{\mathcal T} &=& -12\,[\dot{\mathcal H}^2+\mathcal H\ddot{\mathcal H}]\,, \\ \label{dif2T}
  \dddot{\mathcal T} &=& -12\,[3\dot{\mathcal H}\ddot{\mathcal H}+\mathcal H\dddot{\mathcal H}]\,, \label{dif3T}
\end{eqnarray}
Considering that the modified Friedmann Eqs. (\ref{eq1}) and (\ref{eq2}) can be rewritten as
\begin{eqnarray}
\mathcal H^2&=&-\frac{1}{12f'(\mathcal T)}\,\big[\mathcal T\Omega_{m}+f(\mathcal T)\big]\,,\label{friedfried} \\
\dot{\mathcal H}&=&\frac{1}{4f'(\mathcal T)}\,\big[\mathcal T\Omega_{m}-4\mathcal H\dot{\mathcal T}f''(\mathcal T)\big]\,,\label{acceacce}
\end{eqnarray}
where $\Omega_m$ represents the dimensionless matter density parameter, one obtains two equations for the values of $f$ and $f''$ at present time in terms of the CS, namely
\begin{eqnarray}
f(\mathcal T_0)&=&6\mathcal H_0^2\,(\Omega_{m0}-2)\,,\label{fT}
\end{eqnarray}
and
\begin{eqnarray}
f''(\mathcal T_0)&=&\frac{1}{6\mathcal H_0^2}\Big[\frac{1}{2}-\frac{3\Omega_{m0}}{4(1+q_0)}\Big]\,,\label{f2T}
\end{eqnarray}

\noindent where we are hereafter making use of an additional constraint on the first derivative, i.e. $f'(\mathcal{T}_0) = 1$, which corresponds
to assuming that any $f(\mathcal T)$ choice has to be consistent with the value of the gravitational constant $G$, as measured in the Solar system.
This experimental bound  does not clash with the recipes of cosmography and leaves unaltered the cosmographic analysis.

Afterwards, differentiating Eqs. (\ref{friedfried}) and (\ref{acceacce}) with respect to $t$ up to the fourth order and evaluating them at present time, the expressions for $f'''$ and $f^{(iv)}$ read
\begin{equation}
    f'''(\mathcal T_0)=\frac{1}{(6\mathcal H_0^2)^{2}}\Big[\frac{3}{4}-\frac{3\Omega_{m0}(3q_0^2+6q_0+j_0+2)}{8(1+q_0)^3}\Big]\,,\label{f3T}
\end{equation}
and
\begin{equation}
\begin{split}
    f^{(iv)}(\mathcal T_0)&=-\frac{1}{(6\mathcal H_0^2)^{3}}\Bigl{\{}\frac{3\Omega_{m0}}{16(1+q_0)^5}\,\Big[s_0(1+q_0)+j_0(6q_0^2 \\
    &+17q_0+3j_0+5) +3q_0(5q_0^3+20q_0^2+29q_0 \\
    &+16)+9\Big]+\frac{15}{8} \Bigr\}\,.\label{f4T}
    \end{split}
\end{equation}

 In the next section we will invert the system of Eqs. (\ref{fT})-(\ref{f4T}) in order to obtain the CS in terms of $f(\mathcal T)$ and its derivatives at present time. The resulting expressions will enable us to rewrite the luminosity distance {\it directly} in terms of $f(\mathcal T)$ and its derivatives. Thus, we will be able to fit the quantities under interest, obtaining the cosmographic constraints.

\section{The numerical analysis from cosmography}\label{constraints}

In this section, we consider a Markov Chain Monte Carlo (MCMC) method to constrain the quantities $\{\mathcal H_0, f(\mathcal T_0),\dots, f^{(iv)}(\mathcal T_0)\}$.
In doing so, we notice that the Monte Carlo analysis works fairly well, since it represents a class of algorithms for sampling from probability distributions based on the construction of Markov chains. The expected quality of the sample tends to improve its accuracy as the number of steps increases. Once the number of steps is determined, the convergence is satisfied within an acceptable error, when the initial conditions are well established (see \cite{altrissimo1} for details).

Therefore, first we have to relate the CS to the derivatives of $f(\mathcal T)$ and then we have to fix the initial conditions as viable bounds inferred from observations.
As first step, we take into account the system of equations (\ref{fT})-(\ref{f4T}) and algebraically invert it, in order to obtain the CS in terms of $f(\mathcal T)$.
The resulting expressions for $\mathcal H_0$ and $q_0$ are

\begin{eqnarray}
   \mathcal H_0^2&=&\frac{1}{6}f(\mathcal T_0)(\Omega_{m0}-2)\,,\label{H0}\\
    q_0&=&-\frac{4+4\,f(\mathcal T_0) f''(\mathcal T_0)-(8-3\Omega_{m0})\Omega_{m0}}{2 \mathcal D(f_0,f''_0,\Omega_{m0})}\,,\label{q0}
\end{eqnarray}

\noindent where $\mathcal D(f_0,f''_0,\Omega_{m0})=2+2f(\mathcal T_0) f''(\mathcal T_0)-\Omega_{m0}$.

Note that, according to Eq. (\ref{fT}), $f(\mathcal T_0)$ is negative, so that Eq. (\ref{H0}) for observationally allowed values of $\Omega_{m0}$ is always real. The expressions for $j_0$ and $s_0$ are reported in the Appendix \ref{appendix}.

Next, we make use of the CS in terms of $f(\mathcal T)$ to rewrite the luminosity distance directly in terms of $\Omega_{m0}, f(\mathcal T_0), f''(\mathcal T_0), f'''(\mathcal T_0)$ and $f^{(iv)}(\mathcal T_0)$. Such an expression can be found in the Appendix \ref{appendix}, Eq. (\ref{sestoordinebis}).
We remark that the absence of any dependence on $f'(\mathcal T_0)$ is due to the fact that $f'(\mathcal{T}_0) = 1$, as emphasized in Sec. \ref{cosmography}.
We refer to  CS in terms of $f(\mathcal T)$ and its derivatives as  $CS(f)$.

\begin{figure}
\begin{center}
\includegraphics[width=1.6in]{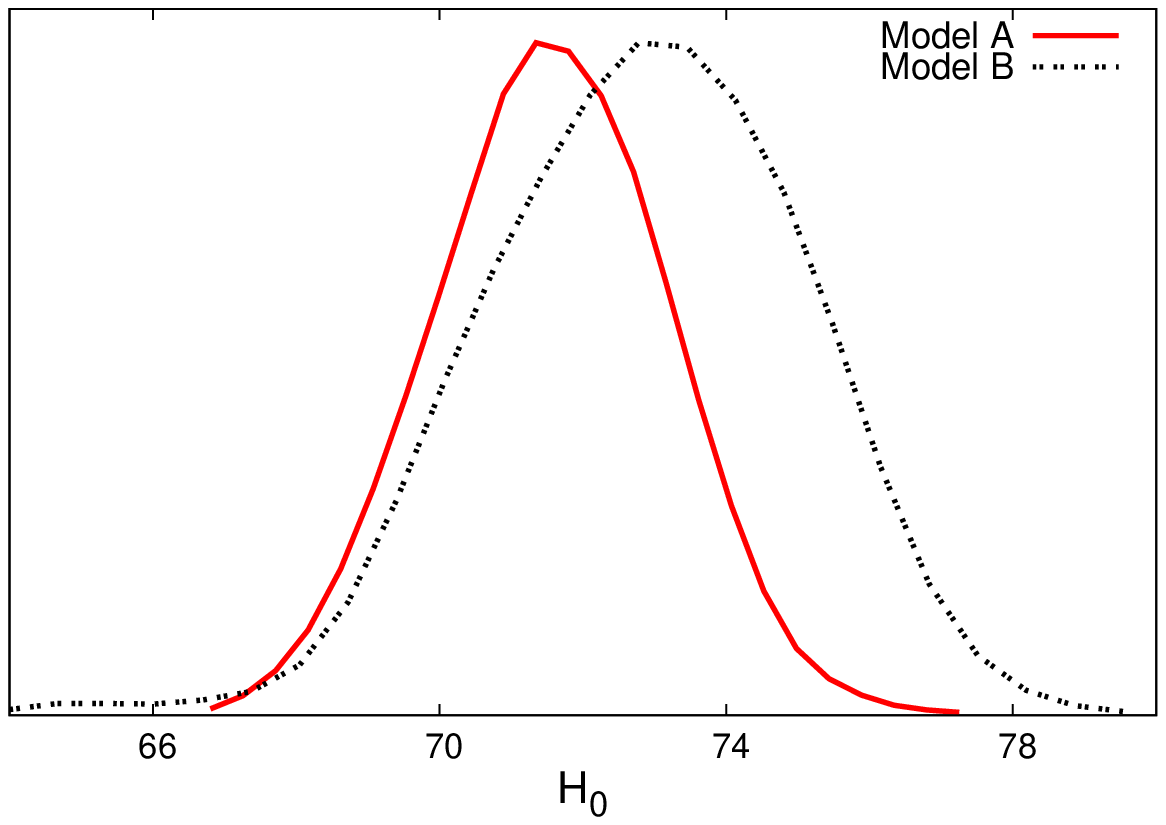}
\includegraphics[width=1.6in]{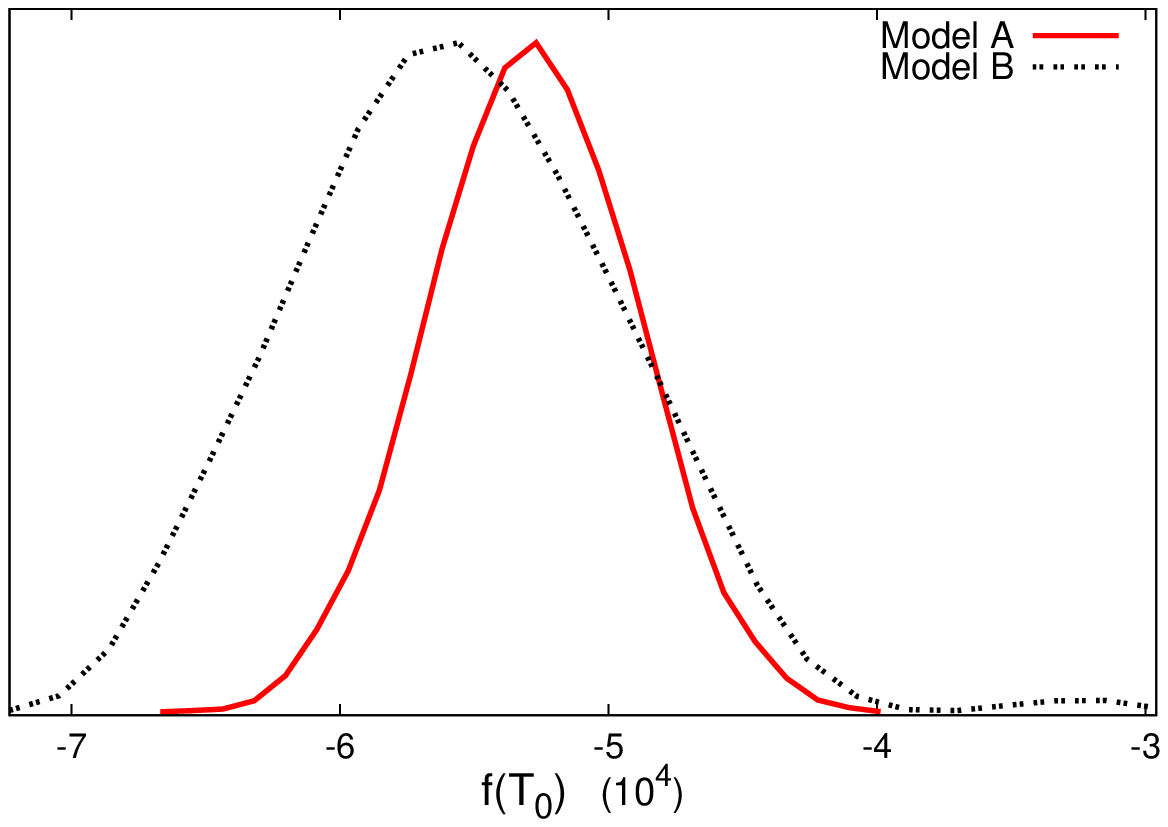}
\includegraphics[width=1.6in]{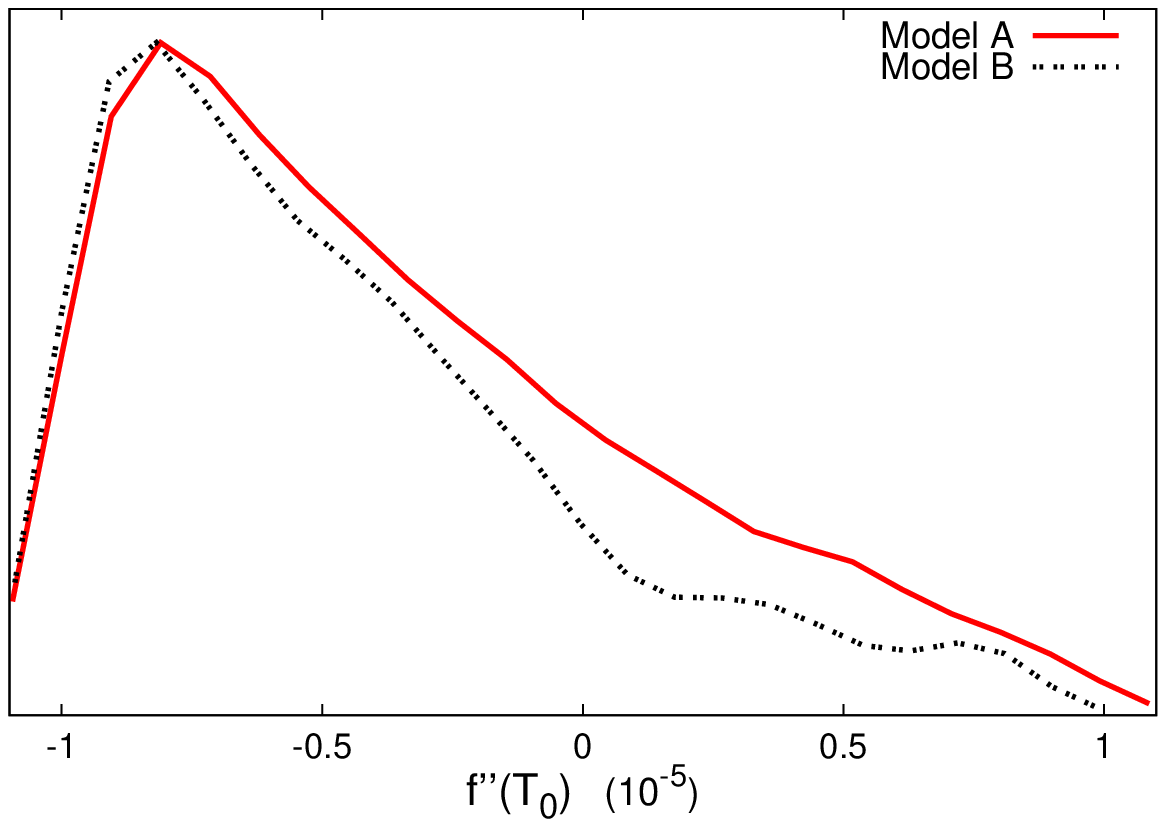}
\includegraphics[width=1.6in]{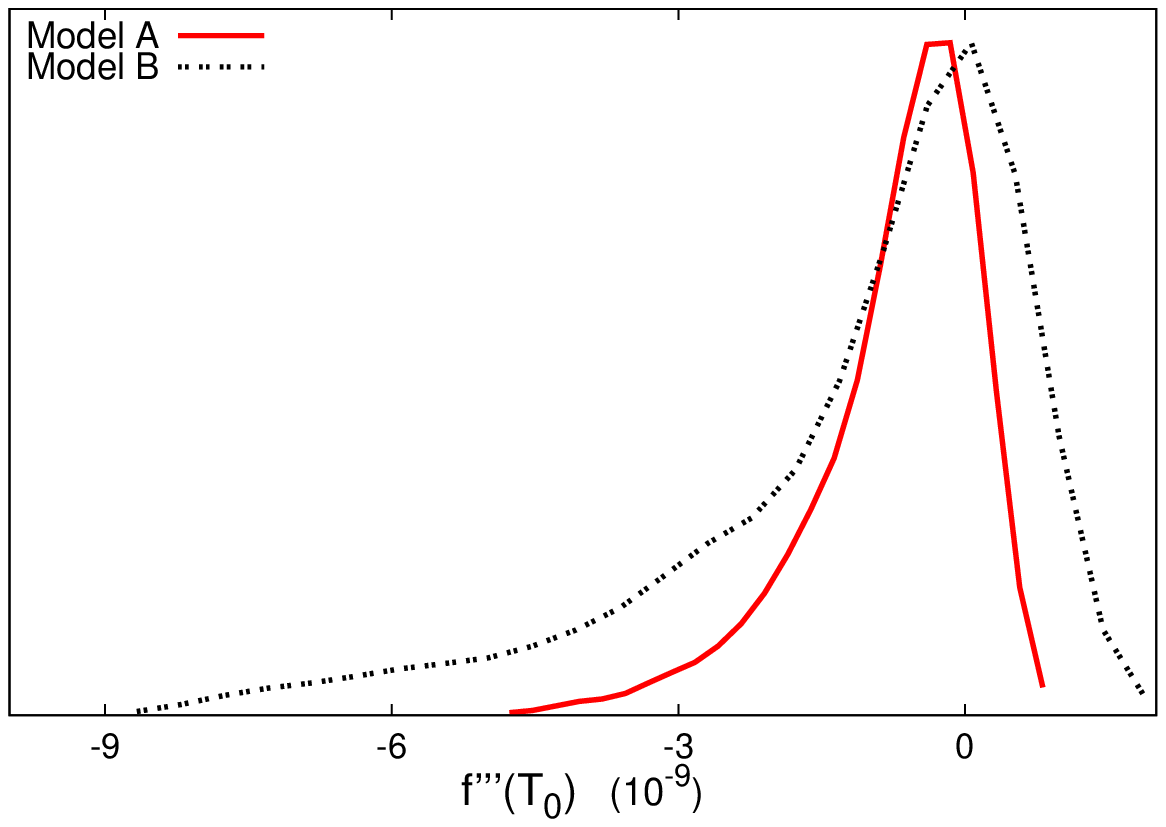}
\includegraphics[width=1.6in]{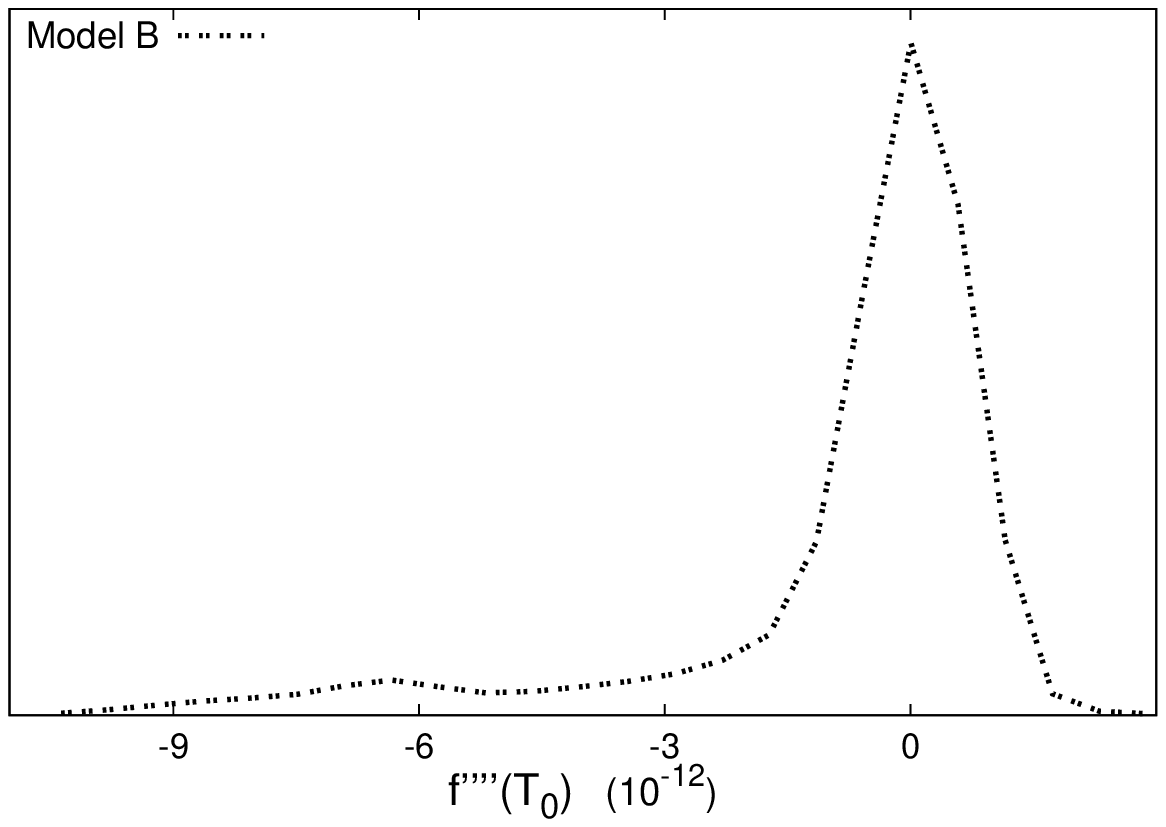}
\caption{1-dimensional marginalized probability for the parameters ex\-plo\-red with MCMC. Solid lines (red) are for model A and dotted lines (black) for model B.}
\end{center}
\label{1dpdf}
\end{figure}

\begin{figure}\label{derived}
\begin{center}
\includegraphics[width=1.6in]{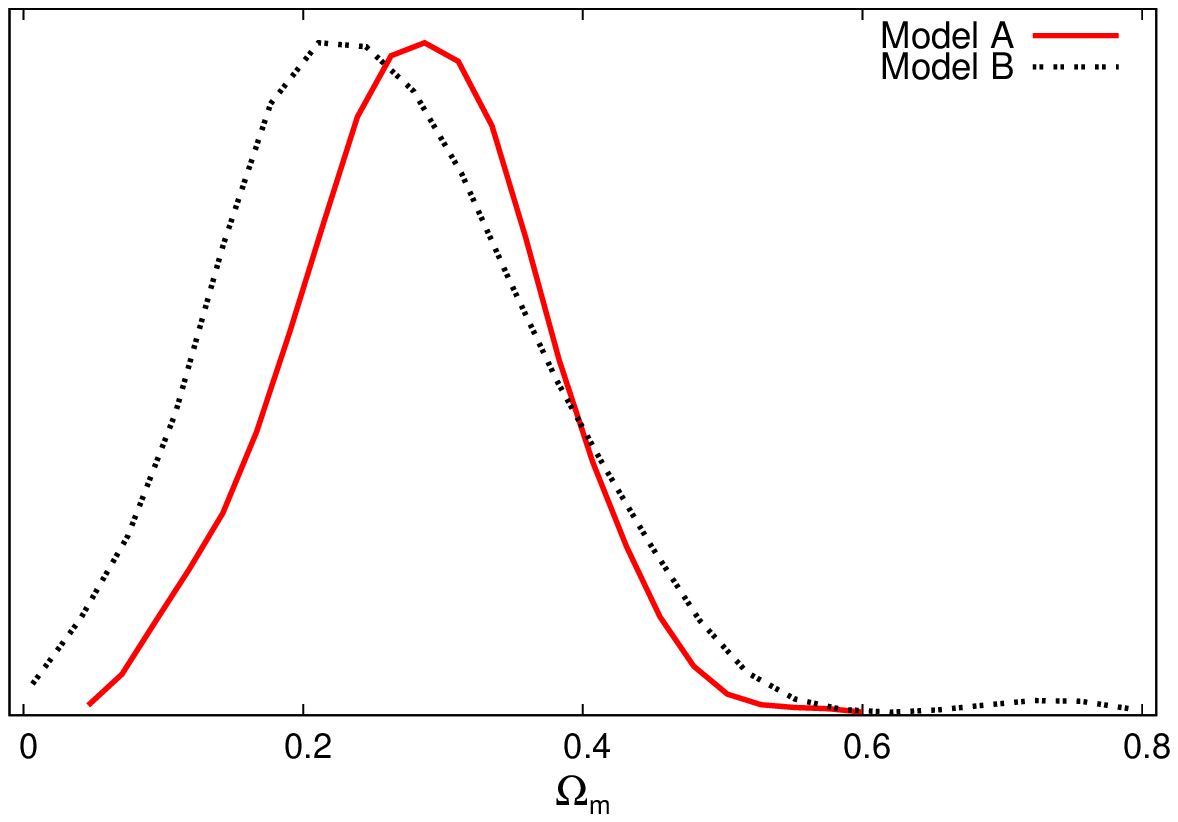}
\includegraphics[width=1.6in]{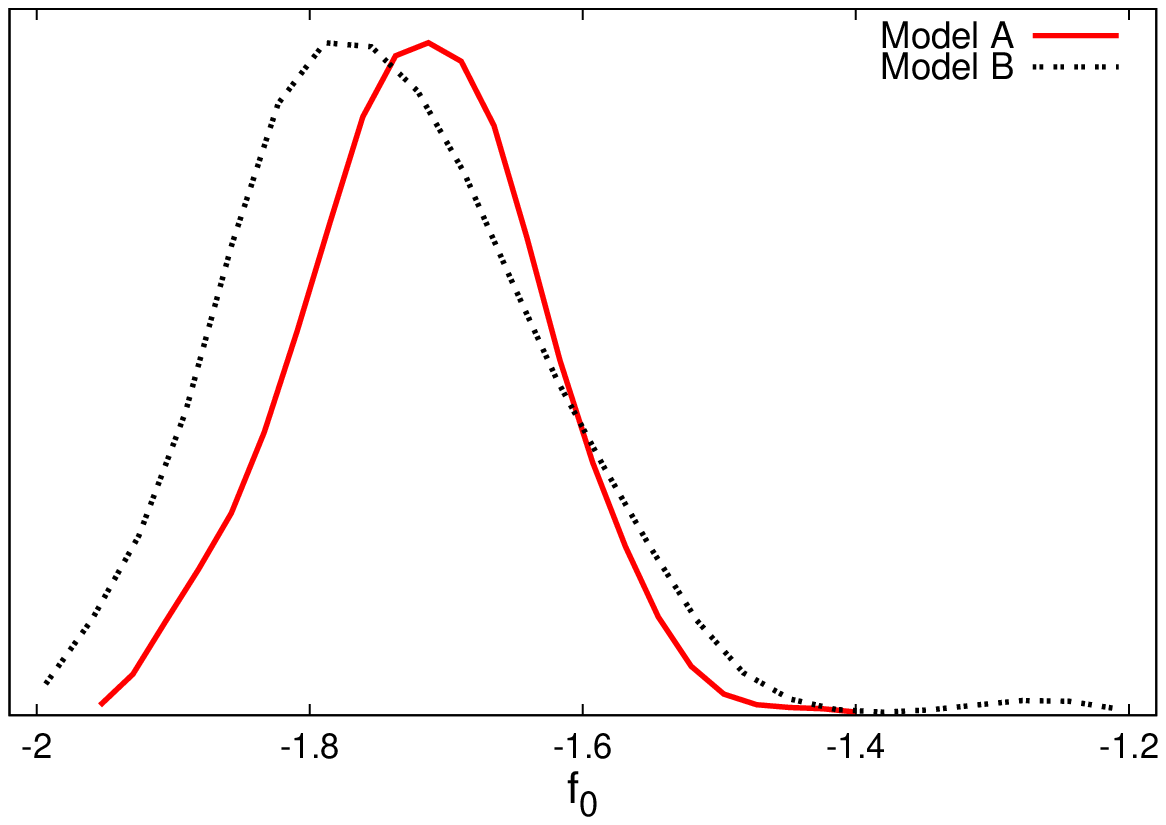}
\includegraphics[width=1.6in]{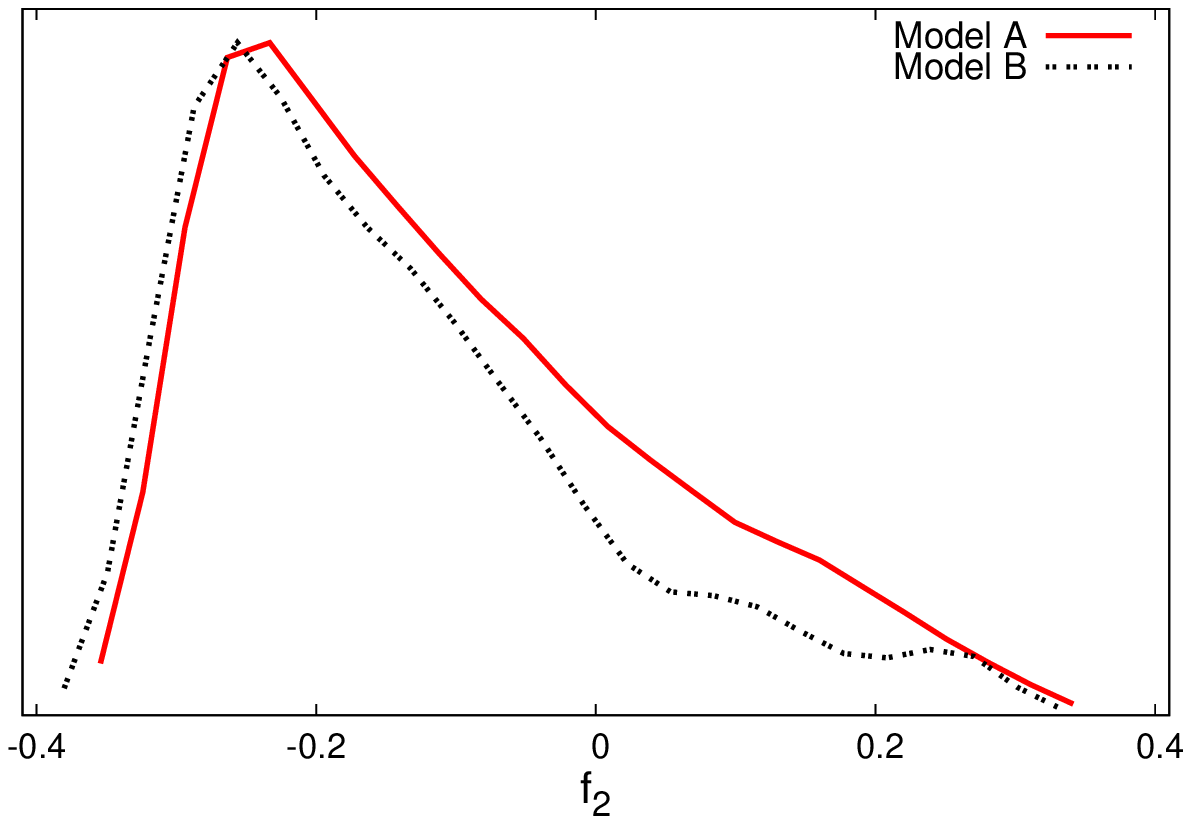}
\includegraphics[width=1.6in]{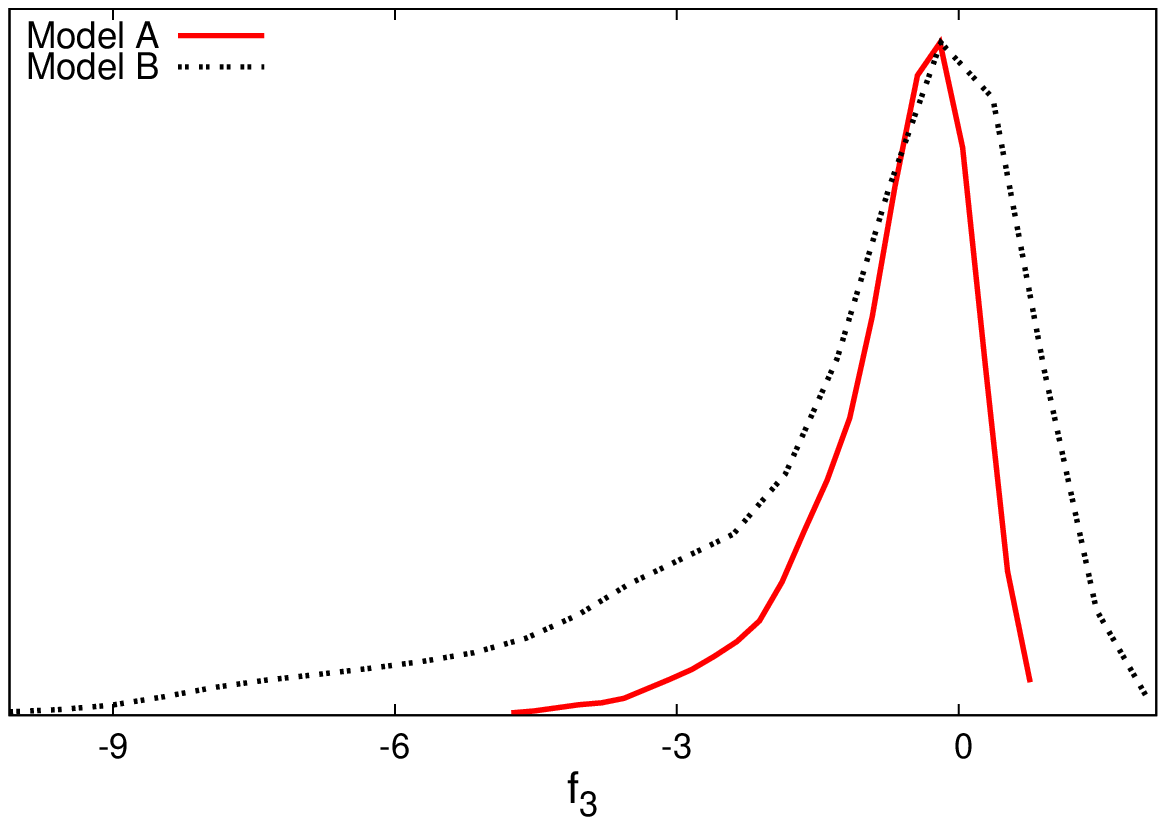}
\includegraphics[width=1.6in]{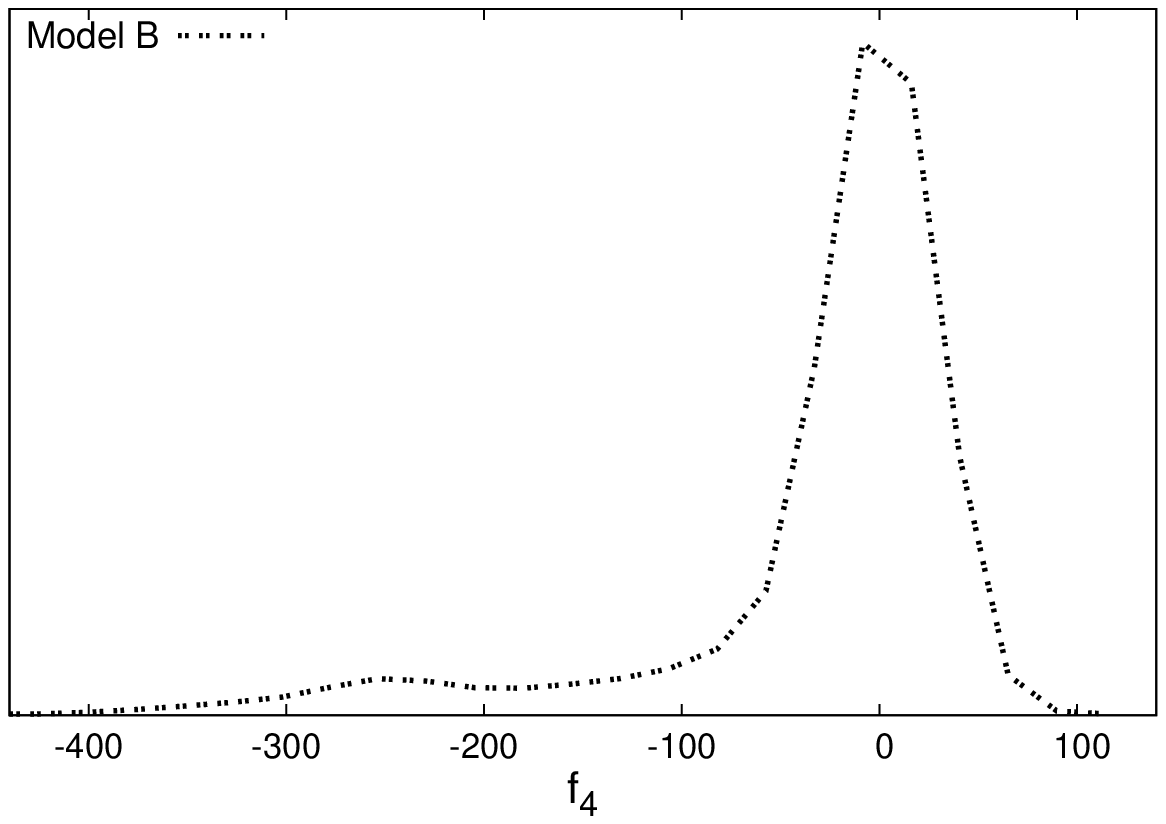}
\caption{1-dimensional marginalized probability for the derived parameters. Solid lines (red) are for model A and dotted lines (black) for model B.}
\end{center}
\end{figure}

\begin{figure}\label{2dpdf}
\begin{center}
\includegraphics[width=1.6in]{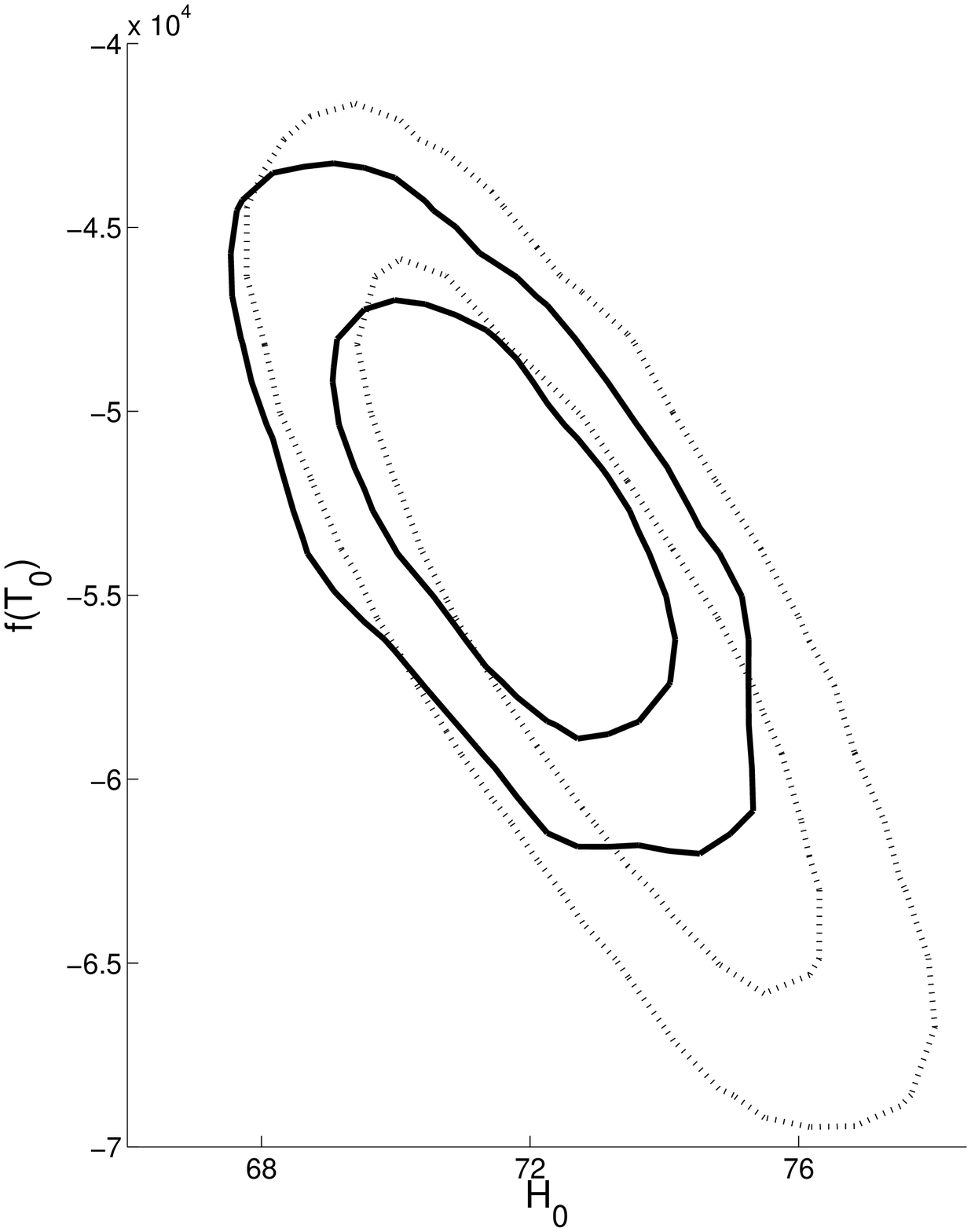}
\includegraphics[width=1.6in]{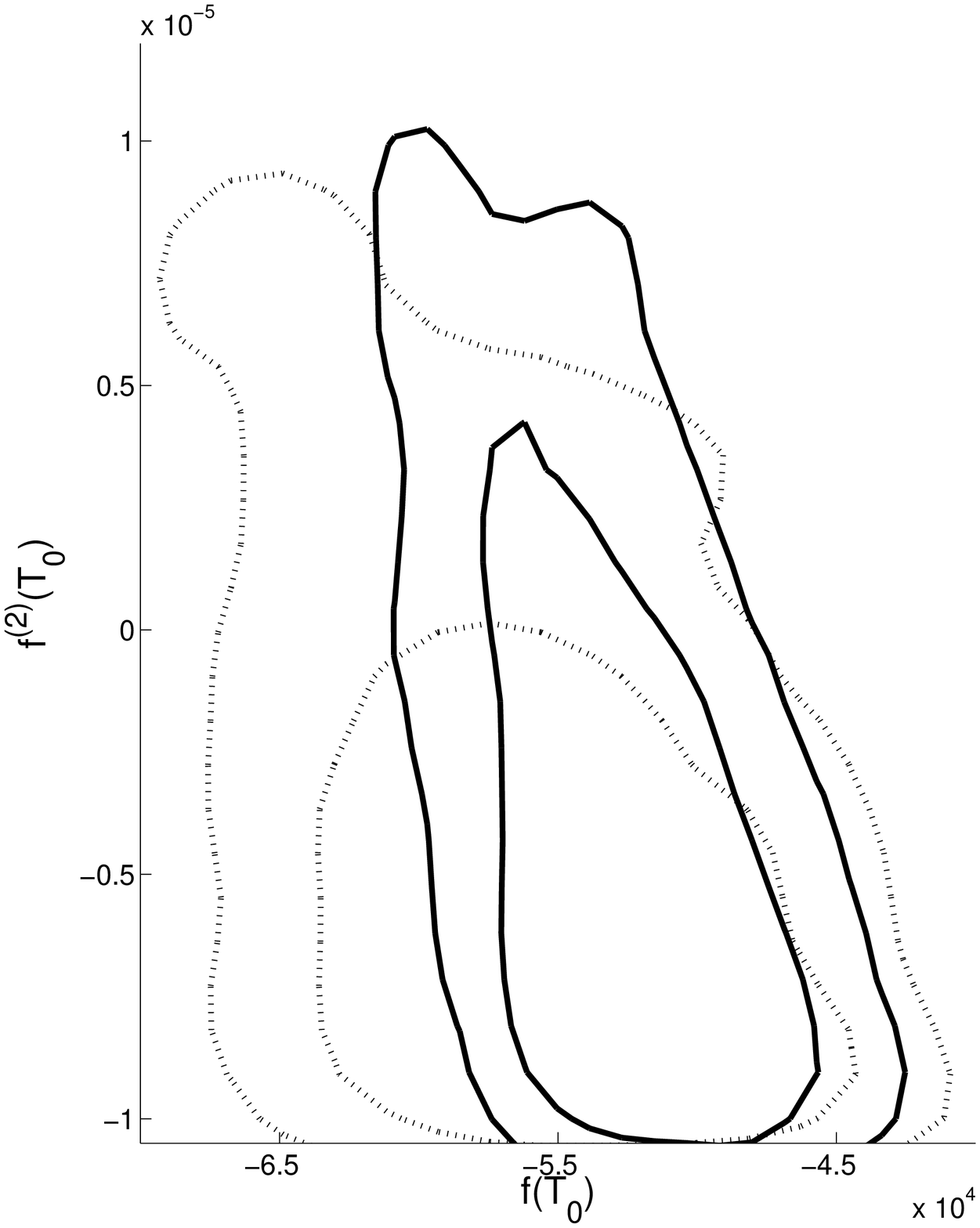}
\includegraphics[width=1.6in]{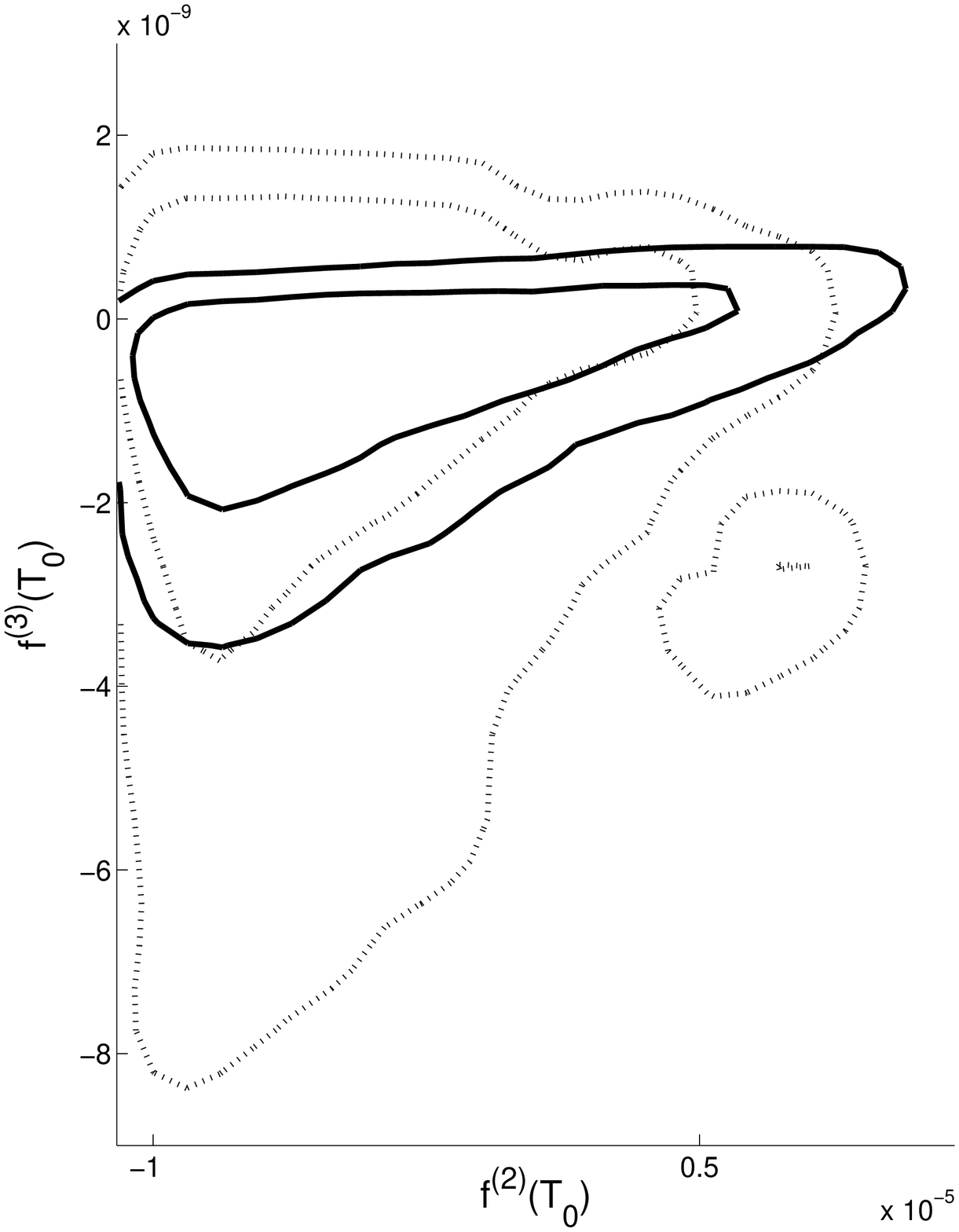}
\includegraphics[width=1.6in]{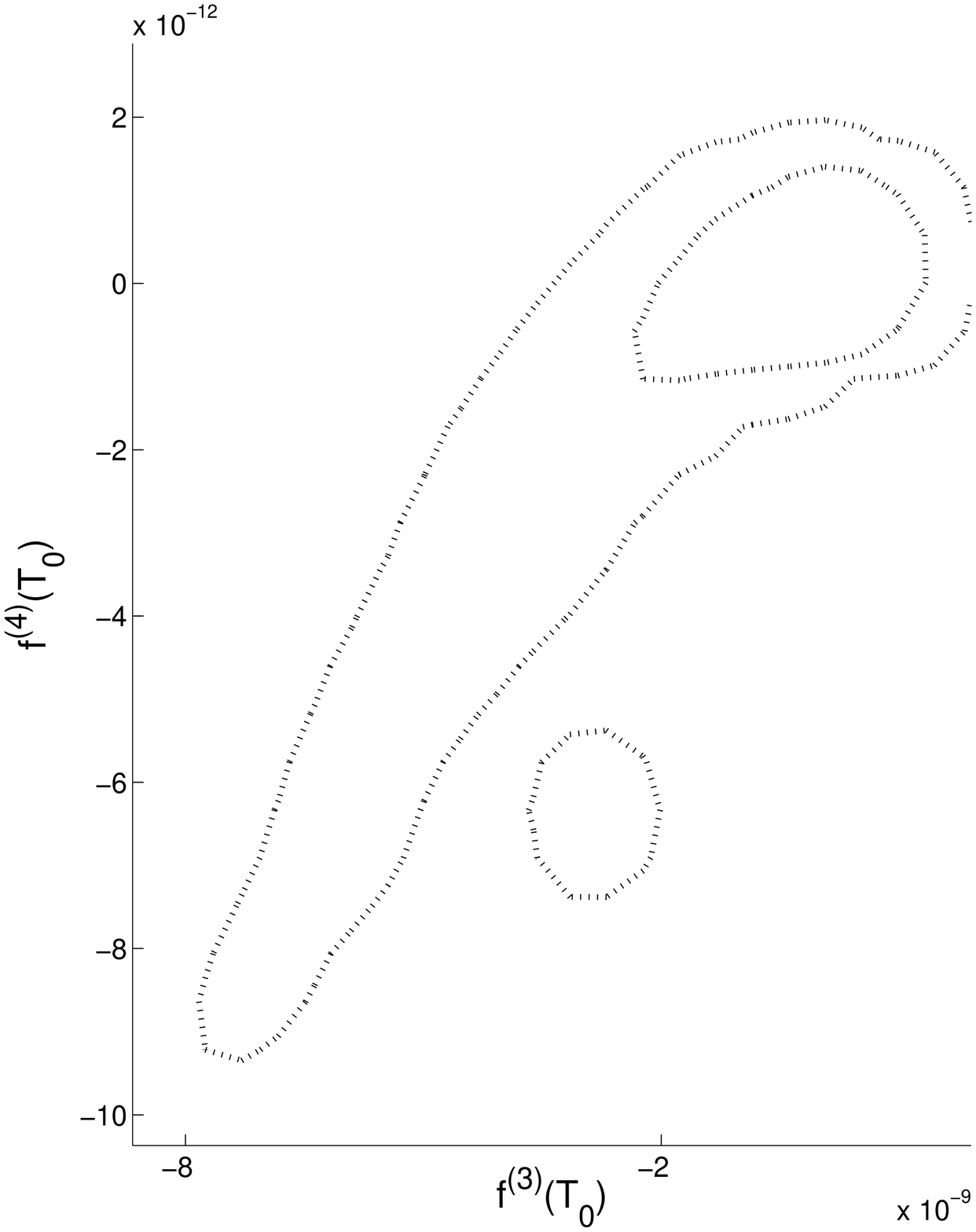}
\caption{2-dimensional marginalized
contours for
consecutive parmeters
 in model A (solid lines)
 and model B (dashed lines).}
\end{center}
\end{figure}

Afterwards, to perform the numerical analysis, we consider two hierarchial classes of models, i.e. $A = \{ \mathcal H_0, f(\mathcal T_0), f''(\mathcal T_0), f'''(\mathcal T_0) \}$ and $B=\{\mathcal H_0, f(\mathcal T_0), f''(\mathcal T_0), f'''(\mathcal T_0), f^{(iv)}(\mathcal T_0)\}$. The corresponding datasets are the Union 2.1 supernovae Ia (SNeIa) compilation \cite{Suzuki:2011hu}, the observational value of the Hubble factor (OHD) in terms of the redshift $z$ and a Gaussian prior on the Hubble constant of $\mathcal H_0 = 74.2 \pm 3.6 \, \text{km/s/Mpc}$ \cite{Riess:2009pu}, which has been extensively measured by the Hubble Space Telescope (HST). The SNeIa sample is an update of the previous compilations  \cite{Amanullah:2010vv} and \cite{Kowalski:2008ez} and includes measurements in the plane $\mu-z$ of 580 supernovae over the redshift range $ 0.015 < z < 1.414$, while we take the compilation of reference \cite{Moresco:2012by} for the OHD measurements, which represents 18 measurements of $\mathcal H(z)$ between the redshift range $ 0.09 < z < 1.75 $.

\begin{widetext}
\begin{table*}[ht]
\caption{\small Table of best fits given by the maximum of the likelihood function. The upper panel contains the parameter spaces ex\-plo\-red with MCMC
for each one of the two models.
The bottom panel contains derived parameters. The data used are the Union 2.1 supernova compilation, observations of the Hubble factor at 18 different redshifts
and a HST measurement of the Hubble constant. The values reported below are in powers of $10^{4}, 10^{-5}, 10^{-9}$ and $10^{-12}$ respectively for $f(\mathcal T_0)$, 
$f''(\mathcal T_0),f'''(\mathcal T_0)$ and $f^{(iv)}(\mathcal T_0)$. The reported error corresponds to the $68 \%$ confidence level.}
\label{table:best_fit}
\centering
\begin{tabular}{c|c|c}
\hline\hline
\hline
$\qquad \qquad$ {\small Parameter} $\qquad \qquad$ & $\qquad \qquad$  {\small Model A}  $\qquad \qquad$  & $\qquad \qquad$  {\small Model B} $\qquad \qquad$ \\ [1.5ex]
                                          & {\small $\chi^2 = 579.8 $ }                           & {\small $\chi^2 = 581.3 $ }   \\[0.8ex]  
\hline
{\small $\mathcal H_0 $}                           & {\small $71.40$}{\tiny${}_{-2.81}^{+2.86}$}           & {\small $71.47$}{\tiny ${}_{-4.17}^{+4.88}$}    \\[0.8ex]

{\small $f(\mathcal T_0)$}           & {\small $-4.932$}{\tiny${}_{-1.205}^{+0.572}$}        & {\small $-5.016$}{\tiny ${}_{-1.709}^{+1.057}$} \\[0.8ex]

{\small $f''(\mathcal T_0)$}          & {\small $-0.963$}{\tiny${}_{-0.037}^{+1.957}$}        & {\small $-0.597$}{\tiny ${}_{-0.403}^{+1.570}$} \\[0.8ex]

{\small $f'''(\mathcal T_0)$}         & {\small $-0.708$}{\tiny${}_{-2.399}^{+1.275}$}        & {\small $0.213$}{\tiny ${}_{-5.389}^{+1.066}$}  \\[0.8ex]

{\small $f^{(iv)}(\mathcal T_0)$}    & {\Large --}                                           & {\small $0.180$}{\tiny ${}_{-4.132}^{+0.818}$}  \\[0.8ex]

\hline

{\small $\Omega_m$}                       & {\small $0.387$}{\tiny${}_{-0.303}^{+0.124}$}          & {\small $0.364$}{\tiny ${}_{-0.327}^{+0.280}$}   \\[0.8ex]  

{\small $f_0$}                            & {\small $-1.613$}{\tiny${}_{-0.303}^{+0.124}$}         & {\small $-1.636$}{\tiny ${}_{-0.327}^{+0.280}$} \\[0.8ex]

{\small $f_2$}                            & {\small $-0.295$}{\tiny${}_{-0.029}^{+0.608}$}         & {\small $-0.183$}{\tiny ${}_{-0.160}^{+0.496}$} \\[0.8ex]

{\small $f_3$}                            & {\small $-0.662$}{\tiny${}_{-2.323}^{+1.155}$}         & {\small $0.201$}{\tiny ${}_{-5.985}^{+1.119}$} \\[0.8ex]

{\small $f_4$}                            & {\Large --}                                            & {\small $5.182$}{\tiny ${}_{-157.65}^{+28.023}$}     \\[0.8ex]

\hline
\end{tabular}

{\small Notes. $\mathcal H_0$ is given in Km/s/Mpc. $\chi^2 \equiv -2 \log L$ is given by the {\it pseudo-chisquared} analysis.}

\end{table*}
\end{widetext}

To constrain the parameters, we use a bayesian method in which the best fits of the parameters are those which maximize the likelihood function
\begin{equation}
 \chi^2 = \chi^2_{\text{Union2.1}} + \chi^2_{\text{HST}} + \chi^2_{\text{OHD}}\,.
\end{equation}
This sum is licit because we are assuming that the different sets of observations are not correlated among them. We therefore perform the Markov Chain Monte Carlo analysis by modifying the publicly available code CosmoMC \cite{Lewis:2002ah}. To obtain the posterior distributions we assume uniform priors over the intervals
\begin{eqnarray}
  10 &<& \mathcal H_0 < 90\,,  \nonumber\\
  -1 &<& 10^{-6}  f(\mathcal T_0) < 1\,, \nonumber\\
  -1 &<& 10^{5}   f''(\mathcal T_0) < 1\,, \\
  -1 &<& 10^{9}  f'''(\mathcal T_0) < 1\,, \nonumber\\
  -2 &<& 10^{12}  f^{(iv)}(\mathcal T_0)< 2\,.\nonumber
\end{eqnarray}

In Tab. \ref{table:best_fit} we show the summary of the constraints. We report the best fit given by the maximum of the likelihood function of the samples, the quoted errors show the $68 \%$ confidence level (c.l.). In Fig. \ref{1dpdf} we plot the corresponding marginalized posterior distributions. For the sake of completeness, it is useful to investigate the dimensionless parameters $f_i$, given by
\begin{eqnarray}
f_i &=& \frac{1}{(6 \mathcal H_0^2)^{-(n-1)}} f^{(i)}(\mathcal T_0)\,,
\end{eqnarray}
whose indexes run over $i=0,2,3,4$. In Fig. \ref{derived} we plot their corresponding marginalized posterior distributions. In the same figure we plot the distribution of the
derived parameter $\Omega_m = f(\mathcal T_0)/6 \mathcal H_0^2 + 2$. The most significant degeneracies are those relating consecutive paramaters. Then, we plot their corresponding two dimensional marginalized
contours at 0.68 c.l. and 0.95 c.l. in Fig. \ref{2dpdf}.  It is important to stress that  combination of parameters do not show considerable degeneracies.

\subsection{An example of a viable $f(\mathcal T)$ model}

The numerical results in Tab. \ref{table:best_fit} favor some classes of $f(\mathcal T)$ and disfavor some others. 
As an example of how to reconstruct a model from the present analysis, we propose a possible example of $f(\mathcal T)$, which we derived by leading to the results of Tab. \ref{table:best_fit}.
We propose the model as follows:
\begin{equation}\label{fTgiusta}
\begin{split}
f(\mathcal T)&=C_0\,\mathcal T+(\mathcal T-\mathcal T_0)\, \Big[
C_1+C_3\,{\rm Cosh}(\mathcal T-\mathcal T_0)\\
&+(\mathcal T-\mathcal T_0)\Big(C_2+C_4(\mathcal T-\mathcal T_0)\,{\rm Sinh}(\mathcal T-\mathcal T_0)\Big)
\Big]\,.
\end{split}
\end{equation}
Equation ($\ref{fTgiusta}$) satisfies the constraints
 $ f(\mathcal T_0)=6\mathcal H_0^2(\Omega_{m0}-2)$
 and
 $ f'(\mathcal T_0)=1$. In addition, we make use of Eq. (\ref{eq: constr}) at present time and we fix
\begin{eqnarray}
  C_0&=&2-\Omega_{m0}\,,\\
  C_1&=&\Omega_{m0}-1-C_3\,.
\end{eqnarray}


By keeping in mind this choice for the first two parameters, the higher order derivatives read
\begin{eqnarray}
f''(\mathcal T_0)&=&2\,C_2\,,\label{2ndorder}\\
f'''(\mathcal T_0)&=&3\,C_3\,,\label{3rdorder}\\
f^{(iv)}(\mathcal T_0)&=&24\,C_4\,.\label{4thorder}
\end{eqnarray}

\noindent Furthermore, using the constraints in Tab. \ref{table:best_fit}, we can set the values
$f''(\mathcal T_0)=-0.6\cdot10^{-5}, f'''(\mathcal T_0)=0.2\cdot10^{-9}$ and $f^{(iv)}(\mathcal T_0)=0.18\cdot10^{-12}$,
giving the numerical values for the free parameters
\begin{eqnarray}
 C_0 &=& 2-\Omega_{m0}\,,\label{C0bis}\\
 C_1&\sim&\Omega_{m0}-1\,,\label{C1bis}\\
 C_2 &=& -3\cdot 10^{-6}\,, \label{C2}\\
 C_3 &=& \frac{1}{15}\cdot10^{-9} \,,\label{C3} \\
 C_4 &=&\frac{3}{4} \cdot 10^{-14}\,.\label{C4}
 \end{eqnarray}

It is clear that the model proposed in Eq. (\ref{fTgiusta}), under the choices of Eqs. (\ref{C0bis})-(\ref{C4}), accurately passes all the numerical bounds presented in this paper and should be candidate for a serious alternative for the $f(\mathcal T)$ class.

\section{Discussion and Conclusions}\label{conclusions}

In this paper, we have described the use of cosmography in order to fix constraints on the values of $f(\mathcal T)$ and its derivatives  up to the fourth order in $\mathcal T$ at present time. In particular, we adopted a procedure which consists in relating the  Cosmographic Set (CS) in terms of $f(\mathcal T)$ and  invert $f(\mathcal T)$ and its derivatives in order to rewrite the luminosity distance in function of such quantities, evaluated at $z=0$. This procedure overcomes the systematical error propagation problems of recent works on cosmography and $f(\mathcal T)$, in which one has to infer the cosmographic bounds once the values of the CS are known. Our procedure allows  to better constrain  the dynamics  and shows that small departures from $\Lambda$CDM seem to be consistent  in the context of $f(\mathcal T)$ theories. The main advantage is that, by keeping in mind the use of cosmography, it is possible to reconstruct the expansion history of $f(\mathcal T)$, without postulating any model \emph{a priori}. Finally,   a possible functional form of $f(\mathcal T)$, compatible with the current cosmographic bounds, can be derived. We expect in future works to improve the present analysis, describing more accurately higher derivatives and giving more insights on the correct form of $f(\mathcal T)$, through the use of different datasets and combining the results with the tests at larger redshift scales.

However, some important remarks are necessary at this point in order to frame the results in a physical context.
For example, we know that the space-time metric (and in particular $a(T)$) is not an observable, but, starting from it, we can derive quantities
related to observables like $H_0$ or $q_0$ that can be connected to the matter-energy and the pressure  by the field equations and experimentally deduced by the luminosity distance $d_L$ or other cosmological indicators. This situation is particularly delicate as soon as one consider alternative theories of gravity like   $f(R)$-gravity or  scalar-tensor theories where one has to select the invariant physical quantities and determine how they behave under conformal and gauge transformations (see \cite{rev2,rev3,bamba} for extended discussions on this topic).  In particular,  one can use conformal transformations to go from the
Jordan frame where matter is minimally coupled to the metric, to the
Einstein frame where the standard Einstein-Hilbert action is present, but the matter
acquires  extra  non-minimal couplings. The conformally transformed
metric is  very different in the two frames (in general, we have a bi-metric structure where a metric is defined in the Einstein frame and the other is defined in the Jordan frame), but they describe the
same physical scenario, and  the observable quantities,
appropriately transformed,  turn out to be the same.
In the present case,   we have  assumed, as discussed in Sec. \ref{cosmography} that we work in the $f(\mathcal{T})$-frame (the Jordan frame). Even though this would suggest to modify the cosmographic quantities, i.e. Eqs. (\ref{pinza}), to guarantee that the conformal transformations hold, it is possible to show that cosmography turns out to be independent in form under  conformal transformations \cite{flanagan,deruelle}. This feature alleviates  the {\it measurement problem} which could influence any analysis. In particular, it prevents measurement departures of cosmological quantities by different observers placed in separate spacetime regions.
To account for such a property, one defines the expressions for $\rho_{\mathcal T}$  and $p_{\mathcal T}$ as in Eqs.(\ref{rhoT}) and (\ref{pT}),  see also
\cite{bamba,capozziellogiafatto}. The results in the Einstein frame, where standard cosmographic series is evaluated, are  restored by considering the further $\rho_{\mathcal T}$  and $p_{\mathcal T}$ terms in the cosmological Friedmann Eqs.(\ref{eq1}) and (\ref{eq2}). In other words, also if the cosmographic quantities appear to be the same, in form, in different frame, the further density and pressure terms make the difference and analogous results are found either in Einstein or in Jordan frame (see the development in Sec.\ref{cosmography}.)

As final remark, it is worth saying that cosmography can be a formidable tool to select physically viable models since its results can be easily translated in different frames. Despite of this achievement,  it is extremely difficult to extend the cosmography results at any redshift due to cosmic evolution and non-reliable data sets at any epoch.

\section*{Acknowledgements}
The authors wants to thank prof. H. Quevedo for discussions and suggestions. We thanks also the anonymous Referee whose suggestions allowed to improve the paper and clarify some delicate points.

\appendix

\begin{widetext}

\section{Luminosity distance in terms of the CS and of the $CS(f)$}\label{appendix}

The expressions for $j_0$ and $s_0$ in terms of $f(\mathcal T)$ and its derivatives when $\mathcal T=\mathcal T_0$, as follows

\begin{eqnarray}
    j_0&=& \frac{1}{2\mathcal D(f_0,f''_0,\Omega_{m0})^3}\Big\{
    16 f(\mathcal T_0)^3 f''(\mathcal T_0)^3
    +(\Omega_{m0}-2)\big[6f(\mathcal T_0)^2(-4f''(\mathcal T_0)^2
    +3f'''(\mathcal T_0)\,\Omega_{m0}^2)\\ \nonumber
    &-&2(\Omega_{m0}-2)^2-3f(\mathcal T_0) f''(\mathcal T_0)(\Omega_{m0}-2)(9\Omega_{m0}^2-4)\big]
    \Big\}\,,\label{j0}
      \end{eqnarray}

\begin{eqnarray}
   s_0&=&\frac{1}{4\mathcal D(f_0,f''_0,\Omega_{m0})^5}\Big\{
   128 f(\mathcal T_0)^5f''(\mathcal T_0)^5
   +2(\Omega_{m0}-2)^5(9\Omega_{m0}-2)
   -8f(\mathcal T_0)^4(\Omega_{m0}-2)\cdot\\\nonumber
   &\cdot&\Big[-81f'''(\mathcal T_0)^2\Omega_{m0}^2
   +90f''(\mathcal T_0)^2f'''(\mathcal T_0)\Omega_{m0}^2
   +4f''(\mathcal T_0)^4
   (10-9\Omega_{m0})
   +27f''(\mathcal T_0)f^{(iv)}(\mathcal T_0)\Omega_{m0}^3
   \Big]\\ \nonumber
   &+&f(\mathcal T_0)f''(\mathcal T_0)(\Omega_{m0}
   -2)^4\Big[40+
   9\Omega_{m0}(9\Omega_{m0}^2+30\Omega_{m0}-16)
   \Big]
   +4f(\mathcal T_0)^3(\Omega_{m0}-2)^2
   \Big[
   36f''(\mathcal T_0)f'''(\mathcal T_0)(5-9\Omega_{m0})\Omega_{m0}^2\\\nonumber
   &+&27f^{(iv)}(\mathcal T_0)\Omega_{m0}^3
   +2f''(\mathcal T_0)^3(40-72\Omega_{m0}-135\Omega_{m0}^2)
   \Big]
   +4f(\mathcal T_0)^2(\Omega_{m0}-2)^3
   \Big[
   -9f'''(\mathcal T_0)^2\Omega_{m0}^2(5+9\Omega_{m0})\\\nonumber
   &+&2f''(\mathcal T_0)^2\big(-20
   +54\Omega_{m0}+27\Omega_{m0}^2(6\Omega_{m0}-5)\big)
   \Big]
   \Big\}\,,\label{s0}
\end{eqnarray}

\noindent where $\mathcal D(f_0,f''_0,\Omega_{m0})=2+2f(\mathcal T_0) f''(\mathcal T_0)-\Omega_{m0}$.

Eqs. (\ref{H0}) and (\ref{q0}), together with (\ref{j0}) and (\ref{s0}) constitute what we call the $CS(f)$, i.e. the algebraic expressions for the cosmographic set in terms of $f(\mathcal T)$ and its derivatives at $z=0$.

Furthermore, the expansion of the luminosity distance $d_L(z)$ in terms of the CS up to the fourth order in $z$, reads
\begin{equation}\label{sestoordine}
\begin{split}
   d_L(z) &=\frac{1}{\mathcal{H}_0} \Bigl[ z +\frac{1}{2} \Bigl(1-
q_0 \Bigr)\,z^2 -\frac{1}{6}
  \Bigl(1-q_0+j_0 -3q_0^2 \Bigr)\,z^3
  +
 \frac{1}{24} \Bigl( 2 + 5 j_0 -
2q_0 + 10 j_0 q_0 -15 q_0^2(1+q_0) +s_0 \Bigr)\,z^4 \\
 &+
\mathcal{O}(z^5)  \Bigr]\,,
\end{split}
\end{equation}

\noindent which is a result evaluated at $k=0$ (i.e. flat FRW cosmology).

It is then easy to plug Eqs. (\ref{H0}), (\ref{q0}), (\ref{j0}) and (\ref{s0}) into (\ref{sestoordine}) to obtain $d_L(z)$ as a function of the $CS(f)$  only. We have

\begin{equation}\label{sestoordinebis}
\begin{split}
   d_L(z) & =  \frac{1}{\mathcal H_0}\Big\{
   z+\frac{8f(\mathcal T_0) f''(\mathcal T_0)+(3\Omega_{m0}-4)(\Omega_{m0}-2)}{4 \mathcal D(f_0,f''_0,\Omega_{m0})}\,z^2
  +\frac{\Omega_{m0}(\Omega_{m0}-2)}{8\mathcal D(f_0,f''_0,\Omega_{m0})^3}\Big[
   40\,\Big(1+f(\mathcal T_0)f''(\mathcal T_0)\Big)^2\\
   &-4\Omega_{m0}\Big(19+28f(\mathcal T_0)f''(\mathcal T_0)+3f(\mathcal T_0)^2f'''(\mathcal T_0)\Big)
   +2\Omega_{m0}^2\Big(23+18f(\mathcal T_0)f''(\mathcal T_0)\Big)-9\Omega_{m0}^3
   \Big]\, z^3\\
   &+\frac{  \Omega_{m0}(\Omega_{m0}-2)}{64\mathcal D(f_0,f''_0,\Omega_{m0})^5}\Big[
 16f(\mathcal T_0)^4\Big(40f''(\mathcal T_0)^2
 -60\Omega_{m0}f''(\mathcal T_0)^2f'''(\mathcal T_0)\Big)+9\Omega_{m0}^2\Big(3f'''(\mathcal T_0)^2-f''(\mathcal T_0)f^{(iv)}(\mathcal T_0)\Big)\\
   &+5(\Omega_{m0}-2)^4\Big(8+9\Omega_{m0}(3\Omega_{m0}-4)\Big)+8f(\mathcal T_0)^3(\Omega_{m0}-2)\Big(
   6\Omega_{m0}f''(\mathcal T_0)f'''(\mathcal T_0)(20
   -33\Omega_{m0})
   +9\Omega_{m0}^2f^{(iv)}(\mathcal T_0)\\
   &+40(9\Omega_{m0}-4)f''(\mathcal T_0)^3
   \Big)-2f(\mathcal T_0)f''(\mathcal T_0)(\Omega_{m0}-2)^3\Big(160+9\Omega_{m0}(57\Omega_{m0}-80)\Big)\\
   &+12f(\mathcal T_0)^2(\Omega_{m0}-2)^2\Big(4f'''(\mathcal T_0)\Omega_{m0}(3\Omega_{m0}-5)
   +f''(\mathcal T_0)^2\big(80+3\Omega_{m0}(69\Omega_{m0}-100)\big)
   \Big)
   \Big]\, z^4+\mathcal{O}(z^5)
   \Big\}\,,
   \end{split}
   \end{equation}
where $\Omega_{m0}$ is given by Eq. (\ref{H0}) and $\mathcal D(f_0,f''_0,\Omega_{m0})=2+2f(\mathcal T_0) f''(\mathcal T_0)-\Omega_{m0}$,
so that $d_L(z)$ is a function of the set $\{\mathcal H_0, f(\mathcal T_0), f''(\mathcal T_0), f'''(\mathcal T_0), f^{(iv)}(\mathcal T_0)\}$ only.
\end{widetext}

\end{document}